\documentclass{elsart}
\newcommand{\be}{\begin{equation}}
\newcommand{\ee}{\end{equation}}
\newcommand{\beq}{\begin{eqnarray}}
\newcommand{\eeq}{\end{eqnarray}}
\newcommand{\mbf}{\mathbf}

\begin{document}

\begin{frontmatter}

\title{Analyzing Stability of Equilibrium Points
in Neural Networks: A General Approach}

\author{Wilson A. Truccolo}

\address{Department of Neuroscience,
Brown University, Providence, RI 02912, USA}
\ead{Wilson\_Truccolo@brown.edu}

\author{Govindan Rangarajan\thanksref{label2}}

\address{Department of Mathematics and Centre for Theoretical Studies,
Indian Institute of Science, Bangalore 560 012, India}
\ead{rangaraj@math.iisc.ernet.in}
\thanks[label2]{Also associated with the Jawaharlal Nehru Centre for
Advanced Scientific Research, Bangalore, India}

\author{Yonghong Chen}

\address{Xi'an Jiaotong University, Xi'an 710049, P.R. China}
\address{Center for Complex Systems and Brain Sciences, Florida Atlantic University, Boca
Raton, FL 33431, USA} \ead{ychen@walt.ccs.fau.edu}

\author{Mingzhou Ding}

\address{Center for Complex Systems and Brain Sciences,
Florida Atlantic University, Boca Raton, FL 33431, USA}
\ead{ding@walt.ccs.fau.edu}

\newpage
\begin{abstract}
Networks of coupled neural systems represent an important class of
models in computational neuroscience. In some applications it is
required that equilibrium points in these networks remain stable
under parameter variations. Here we present a general methodology
to yield explicit constraints on the coupling strengths to ensure
the stability of the equilibrium point. Two models of coupled
excitatory-inhibitory oscillators are used to illustrate the
approach.
\end{abstract}

\end{frontmatter}

\newpage
\section{Introduction}

We consider neural networks of the form \beq\label{geneq1}
\mathbf{\dot{x}}^{i} & = & \mathbf{F}(\mathbf{x}^{i})+\sum_{j=1}^N
G_{ij} \mathbf{H}(\mathbf{x}^j),  \eeq where $\mathbf{x}^{i}$ is
the $M$-dimensional state vector of the $i$th node. Each node can
either be a single neuron ($M=1$) as in Hopfield types of models
\cite{hirsch,wersing}, or a group of neurons ($M>1$), representing
e.g. the cortical column of interacting excitatory and inhibitory
neurons \cite{li3,ding,wilson}. The dynamics of the individual
node is given by $\mathbf{\dot{x}}^{i}
=\mathbf{F}(\mathbf{x}^{i})$ and $\mathbf{H}: R^M\rightarrow R^M $
is the coupling function. The coupling matrix is ${\mbf
G}=[G_{ij}]$ where $G_{ij}$ gives the coupling strength from node
$j$ to node $i$.

Without loss of generality assume that the origin is a stable
equilibrium point for the individual node and remains an
equilibrium point for the network. The stability of the origin
under coupling strength variations is the main concern of the
present work. This problem is mainly motivated by some
computational considerations. For example, a class of models
assert that the background state of the network, represented by
the equilibrium point at the origin, should be quiescent in the
absence of input
\cite{baird,bumann,carpenter,destexhe,li,li1,li2,whittle,yao}.
External inputs, treated as a slowly increasing and then
decreasing function of time, can lead the network through a Hopf
bifurcation to an oscillatory state and then return it to its
background or equilibrium state once the input has been removed.
This natural reset mechanism, requiring the origin to be a stable
equilibrium point, makes the network ready for the next
computational cycle. To endow the oscillatory network the ability
to differentiate patterns of inputs, statistical learning takes
place wherein the coupling strengths between the network units
change according to certain learning rules. Without careful
consideration the learning related parameter changes can
potentially alter the stability of the background state, thereby
defeating the computational picture established earlier. It is
thus desirable to have constraints on the individual coupling
strengths that can be incorporated into the learning rules so that
the stability of the equilibrium point is ensured for all time.

Previous work on stability constraints have mainly concentrated on
recurrent networks of the Hopfield type
\cite{atiya,chen,fang,feng,guez,hirsch,hoppen,kelly,liang,matsuoka,michel,sugawara,wersing,yang}
with $M=1$. In this paper we consider a general approach that
leads to stability bounds on the individual coupling strengths in
recurrent networks with more complex local dynamics. Two explicit
models of coupled neural populations will be used to illustrate
our approach.

\section{Theory}

Our approach consists of three steps.

{\it Step 1}. For simplicity, let
$\mathbf{F}(\mathbf{0})=\mathbf{0},
\mathbf{H}(\mathbf{0})=\mathbf{0}$, and the real parts of the
eigenvalues of the Jacobian $D\mathbf{F}(\mathbf{0})$ be negative
so that the origin is stable for the individual node.

Linearizing Eq. (\ref{geneq1}) around the origin gives (in matrix
form) \beq\label{geneq2} \mathbf{\dot{S}} & = & D\mathbf{F}\cdot
\mathbf{S}+ D\mathbf{H}\cdot \mathbf{S}\cdot \mathbf{G}^T,  \eeq
where
$\mathbf{S}=(\mathbf{x}^1,\mathbf{x}^2,\cdots,\mathbf{x}^N)$.
According to the Jordan canonical form theory, the stability of
Eq. (\ref{geneq2}) is determined by the eigenvalue $\lambda$ of
${\mbf G}$. Let the corresponding eigenvector from ${\mbf G}^T$ be
${\mbf e}$ and let $\mathbf{u}=\mathbf{S}\mathbf{e}$. The
equation for $\mathbf{u}$ reads \be\label{geneq3}
\mathbf{\dot{u}}  = [D\mathbf{F}+\lambda \cdot
D\mathbf{H}]\mathbf{u}. \ee The origin of Eq. (\ref{geneq1}) is
stable if this equation is stable for all the eigenvalues of
${\mbf G}$. This is true even when the coupling matrix is
defective \cite{hirsch1}.

{\it Step 2}. To proceed further we treat $\lambda$ in Eq.
(\ref{geneq3}) as a complex control parameter. Denote by $\Omega$
the region in the Re($\lambda$)- Im($\lambda$) plane where all the
eigenvalues of $(D\mathbf{F}+\lambda\cdot D \mathbf{H})$ have
negative real parts. Clearly, the equilibrium point is stable if
all eigenvalues of $\mbf G$ lie within $\Omega$. We henceforth
refer to $\Omega$ as the stability zone. A schematic of $\Omega$
is shown in Figure 1. We note that $\Omega$ is usually obtained
numerically. For some situations analytical results are possible
(see below).

{\it Step 3}. Thus far the stability criteria are stated in terms
of the eigenvalue of $\mbf G$. The goal in this work is to
directly constraint the coupling strengths themselves. This is
done by making use of the Gershg\"{o}rin disc theorem \cite{horn}.

Given an $n \times n$ matrix ${\mbf A}=[a_{ij}]$, the
Gershg\"{o}rin theorem states that all eigenvalues of $\mbf A$ are
located in the union of $n$ discs (called the Gershg\"{o}rin
discs) where each disc is given by \beq
 \{ z \in C : |z-a_{ii}| \leq \sum_{j \neq i} |a_{ji}| \}, \ \ \
 i=1,2, \ldots n. \nonumber
\eeq Alternative forms of the $n$ discs are \cite{horn}: \beq
 \{ z \in C : |z-a_{ii}| \leq \sum_{j \neq i} |a_{ij}| \}, \ \ \
 i=1,2, \ldots n. \nonumber
\eeq Combining the two, we have the form used in the remainder of
this paper: \beq
 \{ z \in C : |z-a_{ii}| \leq \frac{1}{2} \sum_{j \neq i}
(|a_{ji}|+|a_{ij}|) \}, \ \ \
 i=1,2, \ldots n.
\eeq This form is more intuitive since it involves incoming and
outgoing coupling strengths for a given node.

The stability conditions for the equilibrium point can now be
stated as follows:

\begin{enumerate}
\item The center $G_{ii}$ ($i=1,2, \ldots , N$) of every
Gershg\"{o}rin disc of $\mbf G$ lies inside the stability zone
$\Omega$;

\item The radius of every Gershg\"{o}rin disc is shorter than the
distance from the center of the disc to the boundary of $\Omega$.
\end{enumerate}

In other words, letting $\delta(x)$ denote the distance from point
$x$ on the real axis to the boundary of $\Omega$, stability of
the equilibrium point is ensured if \beq (G_{ii},0) \in \Omega
\;\;\; \mbox{and} \;\;\;\; \frac{1}{2}{\sum_{j \neq
i}}(|G_{ji}|+|G_{ij}|) < \delta(G_{ii}) \label{gnrlstbl} \eeq for
$i=1,2, \ldots, N$.

\section{Examples}

\subsection{The case of $M=1$}

When one dimensional systems are coupled together, the matrices
$D\mbf F$ and $D \mbf H$ are reduced to real numbers.
Representing them by $\mu$ and $\nu$ respectively, the stability
zone is easily obtained as Re$(\lambda)<-\mu/\nu$. The distance
from the center of the $i$th Gershg\"{o}rin disc to the boundary
of $\Omega$ is given by $\delta(G_{ii}) =-\mu/\nu-G_{ii}$. Using
Eq. (\ref{gnrlstbl}) we obtain the stability conditions as \be
\frac{1}{2}{\sum_{j \neq i}}(|G_{ji}|+|G_{ij}|)+G_{ii} <
-\mu/\nu. \ee This result was obtained before in
\cite{hirsch,wersing}.

\subsection{A coupled oscillator model with $M=2$}

The general topology for the model is shown in Figure 2. The basic
unit in the model is a neural population consisting of either
excitatory or inhibitory cells \cite{baird,li,li4,wilson}. The
functional unit in the network is a cortical column consisting of
mutually coupled excitatory and inhibitory populations. The
columns are then coupled through mutually excitatory interactions
to form the network.

A single column is described by two first order differential
equations

\parbox{10cm}{\beq\label{eieq1} \frac{d x}{d t} + a x & = & - k_{ei}
Q(y,Q_m) + I, \nonumber \\ \frac{d y}{d t} + b y & = & k_{ie}
Q(x,Q_m). \nonumber \eeq} \hfill
\parbox{1cm}{\beq  \eeq}

Here $x$, $y$ represent the local field potentials of the
excitatory and inhibitory populations, respectively, and $I$ is
the input ($I=0$ in the subsequent analysis). The constants
$a,b>0$ are the damping constants. The parameter $k_{ie}>0$ gives
the coupling gain from the excitatory ($x$) to the inhibitory
($y$) population whereas $k_{ei} > 0$ represents the strength of
the reciprocal coupling. The nonlinear neuronal interaction is
realized through the sigmoid function $Q(\cdot,Q_m)$ where $Q_m$
is a parameter controlling the slope of the function. Here we only
need to specify that $Q(0,Q_m)=0$ and $Q'(0,Q_m)=1$.

The $N$ columns are coupled together in the following fashion:

\parbox{10cm}{\beq\label{neteqns} \frac{d x_n}{d t} + a x_n & = & - k_{ei}
Q(y_n,Q_m) + \frac{1}{N} \sum_{p=1}^N c_{np} Q(x_p,Q_m) + I_n,
\nonumber
\\ \frac{d y_n}{d t} + b y_n & = & k_{ie} Q(x_n,Q_m), \nonumber
\eeq} \hfill \parbox{1cm}{\beq \eeq}

 where the columns are indexed
by $n=1,2, \ldots , N$ and the coupling strength $c_{np}$ is the
gain from the excitatory population of column $p$ to the
excitatory population of column $n$.

Variables used in Eq. (\ref{geneq3}) can be explicitly evaluated
for the present model as \beq D \mbf F= \pmatrix{ -a & -k_{ei} \cr
k_{ie} & -b},\quad [\mathbf{G}]_{np} = \frac{c_{np}}{N},\quad D
\mbf H = \pmatrix{1 & 0 \cr 0 & 0}. \nonumber \eeq where we have
used the fact $Q'(0,Q_m)=1$.

To discover the stability zone we study the eigenvalue $\alpha$ of
the matrix $({D \mbf F}+\lambda \cdot{D \mbf H})$ as a function of
$\lambda$. The characteristic polynomial of this matrix is given
by \beq f(\alpha)=\alpha^2 + \alpha (a+b-\lambda) + (k_{ei} k_{ie}
+ ab -b \lambda). \nonumber \eeq For an arbitrary coupling matrix
${\mbf G}$, its eigenvalues $\lambda$ could be complex: \beq
\lambda = \lambda_R + i \lambda_I.  \nonumber \eeq Then the
characteristic polynomial becomes \beq f(\alpha) = \alpha^2 +
\alpha (a+b-\lambda_R-i \lambda_I) + (k_{ei} k_{ie} + ab -b
\lambda_R - i b \lambda_I).\nonumber \eeq

The range of parameter values which gives $\mathrm{Re}(\alpha) <
0$ can be determined by applying the generalized Routh-Hurwitz
criterion (see Appendix I). Following this procedure, consider $-i
f(i \alpha)$: \beq -i f(i \alpha) = i \alpha^2 + \alpha
(a+b-\lambda_R) -i \alpha \lambda_I - i (k_{ei} k_{ie} + ab -b
\lambda_R) -  b \lambda_I. \nonumber \eeq This has to be put into
the following standard form: \beq -i f(i \alpha) = b_0 \alpha^2 +
b_1 \alpha + b_2 + i [a_0 \alpha^2 + a_1 \alpha + a_2 ]. \nonumber
\eeq Comparing the two equations we get \beq
a_0 & = & 1, \ \ a_1 = - \lambda_I, \ \ a_2 = - (k_{ei} k_{ie} + ab -b \lambda_R), \nonumber \\
b_0 & = & 0, \ \ b_1 = (a+b-\lambda_R), \ \ b_2 = -  b \lambda_I.
\nonumber \eeq

Applying the generalized Routh-Hurwitz criterion, we have
$\mathrm{Re}(\alpha) < 0$ if the following two conditions are met:
\beq \nabla_2 = \vline
\begin{array}{cc}
1 & - \lambda_I \\
0 & (a+b-\lambda_R)
\end{array}
\vline > 0 \nonumber \eeq and \beq \nabla_4 = \vline
\begin{array}{cccc}
1 & - \lambda_I & - (k_{ei} k_{ie} + ab -b \lambda_R) & 0 \\
0 & (a+b-\lambda_R) & - b \lambda_I & 0 \\
0 & 1 & - \lambda_I & - (k_{ei} k_{ie} + ab -b \lambda_R) \\
0 & 0 & (a+b-\lambda_R) & - b \lambda_I
\end{array}
\vline > 0.  \nonumber \eeq

Evaluating the above determinants and simplifying, we get

\parbox{10cm}{\beq\label{nonsymineqs}
(a+b-\lambda_R) & > & 0, \nonumber \\
(k_{ei} k_{ie} + ab -b \lambda_R) (a+b-\lambda_R)^2 - b
\lambda_I^2 (\lambda_R-a) & > & 0. \nonumber \eeq} \hfill
\parbox{1cm}{\beq \eeq}

Solving the inequalities, the stability zone $\Omega$ (see Figure
3) is found to be the region to the left of the curve \beq
\lambda_I^2=\frac{(k_{ei} k_{ie} + ab -b \lambda_R)
(a+b-\lambda_R)^2} {b(\lambda_R-a)} . \label{bdryeqs}\eeq The
pointed tip of the curve in Figure 3 along the real axis is given
by $({\rm min}(a+b,a+k_{ie}k_{ei}/b),0)$ and it corresponds to the
symmetric coupling case.

The distance $\delta(G_{ii})$ from the center of the $i$th
Gershg\"{o}rin disc to the boundary is (see Appendix II for more
details) \beq \label{short}
\delta(G_{ii})=\sqrt{(a-G_{ii})^2-b^2-2k_{ie}k_{ei}+
2\sqrt{k_{ie}k_{ei}[2b(a+b-G_{ii})+k_{ie}k_{ei}]}}. \nonumber \eeq
So the stability conditions [cf. Eq.(\ref{gnrlstbl})] are given by
\beq G_{ii} & < & {\rm min}(a+b,a+k_{ie}k_{ei}/b), \nonumber \\
\frac{1}{2}{\sum_{j \neq i}} & ( & |G_{ji}| + |G_{ij}|) < \nonumber \\
& & \sqrt{(a-G_{ii})^2-b^2 -
2k_{ie}k_{ei}+2\sqrt{k_{ie}k_{ei}[2b(a+b-G_{ii})+k_{ie}k_{ei}]}}.
\nonumber \eeq

We note that, since the boundary curve of the stability zone
asymptotically approaches the straight line $\lambda_R=a$, we can
use this line to define a new stability zone to obtain some
simpler stability constraints. The distance to the new boundary is
easily found to be \beq \delta_{i}=|a-G_{ii}|. \nonumber \eeq In
this case, the stability condition simplifies to \beq
\frac{1}{2}{\sum_{j \neq i}}(|G_{ji}|+|G_{ij}|)+G_{ii} < a, \quad
i=1,2, \ldots , N. \label{simple} \eeq This simplified condition
is a good approximation if ${\rm min}(a+b,a+k_{ie}k_{ei}/b)$ is
sufficiently close to $a$. We further note that Eq. (\ref{simple})
is satisfied if \beq |G_{ij}|< a/N,\quad i,j = 1,2, \ldots , N.
\nonumber \eeq That is, the equilibrium point is stable if \beq
|c_{np}|< a,\quad \forall \ n,p = 1,2, \ldots , N. \nonumber \eeq
This simple stability bound on the individual coupling strengths
can be very useful in practice.

\subsection{A coupled oscillator model with $M=4$} The previous model
represents a neural population by a first order differential
equation. This has the property that its impulse response has a
instantaneous rise phase. Here we consider another model where the
neural population is a second order differential equation
possessing a finite rise and decay impulse response. Each
individual column is described by a system of two second order
differential equations \cite{freeman}:

 \parbox{10cm}{\beq \label{eieq2} \frac{d^2 x}{d t^2} + (a+b)
\frac{d x}{d t} + a b x & = & - k_{ei}
 Q(y,Q_m) + I, \nonumber \\
\frac{d^2 y}{d t^2} + (a+b) \frac{d y}{d t} + a b  y & = & k_{ie}
Q(x,Q_m). \nonumber \eeq} \hfill  \parbox{1cm} {\beq \eeq}

The parameters have the same interpretation as before. The $N$
column equations are given by: \newpage \beq\label{neteqns2}
\frac{d^2 x_n}{d t^2} + (a+b) \frac{d x_n}{d t} + a b x_n & = & -
k_{ei} Q(y_n,Q_m) +
\nonumber \\& & \frac{1}{N} \sum_{p=1}^N c_{np} Q(x_p,Q_m) + I_n, \\
\frac{d^2 y_n}{d t^2} + (a+b) \frac{d y_n}{d t} + a b  y_n & = &
k_{ie} Q(x_n,Q_m), \nonumber \eeq where the same network topology
in Figure 2 applies.

We first consider the stability of the single column equations
given in Eq. (\ref{eieq2}). When the input $I$ is zero, the
origin $x=0,y=0$ is an equilibrium point. In order to study its
stability properties, we convert the above second order
differential equations to the following system of first order
differential equations: \beq
\frac{d z_1}{d t} & = & z_2, \nonumber \\
\frac{d z_2}{d t} & = & -(a+b) z_2 - a b z_1 - k_{ei} Q(z_3,Q_m), \nonumber \\
\frac{d z_3}{d t} & = & z_4, \nonumber \\
\frac{d z_4}{d t} & = & -(a+b) z_4 - a b z_3 + k_{ie} Q(z_1,Q_m),
\nonumber \label{syseqns2} \eeq where \beq z_1 = x, \ \ z_2 =
\frac{d x}{d t}, \ \ z_3 = y, \ \ z_4 = \frac{d y}{d t}. \nonumber
\eeq

The Jacobian matrix $D{\mbf F}$ is obtained as \beq D{\mbf F}=
\pmatrix{ 0 & 1 & 0 & 0 \cr - a b & -(a+b) & -k_{ei} & 0 \cr 0 & 0
& 0 & 1 \cr k_{ie} & 0 & - a b & -(a+b)}. \label{A2}\eeq Here we
have used the fact that $Q'(0,Q_m)=1$. For stability of the
origin, the real parts of all eigenvalues of $D{\mbf F}$ should be
less than zero. The eigenvalues are determined from the
characteristic equation: \beq \lambda^4+2(a+b)
\lambda^3+(a^2+4ab+b^2) \lambda^2 + 2(a^2b+ab^2) \lambda + k_{ie}
k_{ei} + a^2 b^2 = 0. \nonumber \eeq Applying the Lienard-Chipart
criterion (see Appendix I), the real parts of all eigenvalues are
negative if the following inequalities be satisfied: \beq
a^2 b^2  +  k_{ie} k_{ei} >  0, \nonumber \\
a^2 b  +  a b^2  >  0, \nonumber \\
a +  b  >  0, \nonumber \\
 -k_{ie}k_{ei} +  ab(a+b)^2
> 0. \nonumber \eeq Since $a,b,k_{ei},k_{ie} > 0$, the first three
inequalities are automatically satisfied. After simplification,
the last inequality can be written as: \beq k_{ie}k_{ei} <
ab(a+b)^2. \label{stabo}\eeq To summarize, the origin is stable
for the single column equations if the above condition is
satisfied. Henceforth, we will assume that this is true.

Next, we consider the stability of a network of coupled columns
given in Eq. (\ref{neteqns2}). Here \beq [\mathbf{G}]_{np} =
\frac{c_{np}}{N}, \nonumber \eeq and \beq {D\mbf H} = \pmatrix{0 &
0 & 0 & 0 \cr 1 & 0 & 0 & 0 \cr 0 & 0 & 0 & 0 \cr 0 & 0 & 0 & 0 }
\nonumber \eeq As before, we examine the eigenvalue $\alpha$ of
the matrix ${D\mbf F}+\lambda \cdot {D \mbf H} $ as a function of
$\lambda$. The characteristic polynomial of this matrix is given
by \beq f(\alpha) & = & \alpha^4 + 2(a+b) \alpha^3 + [(a+b)^2 + 2
a b
-\lambda]\alpha^2   \nonumber \\
& & +[2 a b (a+b) - \lambda (a+b)] \alpha + [a^2 b^2 - a b \lambda
+ k_{ie} k_{ei}]. \nonumber \eeq

For complex $\lambda$, we are not able to obtain an analytical
form for the stability zone $\Omega$, since the characteristic
equation results in a 8th order polynomial when applying the
generalized Routh-Hurwitz criterion. However, numerical results
are always possible. Figure 4 shows the stability zone $\Omega$
when $a=0.22, b=0.72,k_{ie}=0.1,k_{ei}=0.4$. After numerically
finding the distance $\delta(G_{ii})$ from the center of the $i$th
Gershg\"{o}rin disc to the boundary curve, Eq. (\ref{gnrlstbl})
can again be used to give the stability criteria.

If the coupling is symmetric, which implies that $\lambda$ is
real, the stability boundary is just the rightmost tip of the
curve along the real axis in Figure 4. Then the distance $\delta$
is given by the absolute difference between the coordinates of
the tip point and the center of the $i$th Gershg\"{o}rin disc.
This tip can be determined as follows.

Again applying the Lienard-Chipart criterion (see Appendix I), the
real parts of all eigenvalues are negative if the following
inequalities are satisfied:

 \parbox{10cm}{\beq \label{ineq2}
a^2 b^2 - a b \lambda + k_{ie} k_{ei} & > & 0, \nonumber \\
2 a b (a+b) - \lambda (a+b) & > & 0, \nonumber \\
(a+b) & > & 0, \nonumber \\
\lambda^2 - 2 (a+b)^2 \lambda + 4(a^3 b + 2 a^2 b^2 + a b^3 -
k_{ie} k_{ei} )& > & 0. \nonumber \eeq} \hfill
\parbox{1cm} {\beq \eeq}

Since $a$, $b$ are positive, the third inequality is automatically
satisfied. After simplification, the first two inequalities
become: \beq
\lambda & < & \frac{k_{ie} k_{ei}+a^2 b^2}{a b}, \nonumber \\
\lambda & < & 2 a b. \nonumber \eeq

The last inequality is of the form \beq a_1 \lambda^2 - a_2
\lambda + a_3 > 0, \nonumber \eeq where \beq a_1 = 1, \ \ a_2 = 2
(a+b)^2, \ \ a_3 = 4[a b (a+b)^2 - k_{ie} k_{ei}]. \nonumber \eeq
Note that $a_1$, $a_2$ are obviously positive. It turns out $a_3$
is also positive because of the local stability condition derived
in Eq. (\ref{stabo}). The quadratic function $a_1 \lambda^2 - a_2
\lambda + a_3$ with $a_1$, $a_2$, $a_3$ positive has a unique
global minimum at $\lambda = a_2/2 a_1$. Thus the minimum occurs
at a positive value of $\lambda$. It is also seen that \beq a_2^2
- 4 a_1 a_3 = 4 \left[ (a+b)^4 - 4 [a b (a+b)^2 - k_{ie} k_{ei}]
\right]. \nonumber \eeq This can be simplified as \beq a_2^2 - 4
a_1 a_3 = 4 \left[ (a^2-b^2)^2 + 4 k_{ie} k_{ei} \right],
\nonumber \eeq which is positive since $k_{ie} k_{ei}$ is
positive. Thus both the zeros of the quadratic function (we will
denote them $\eta_1$ and $\eta_2$ with $\eta_1 < \eta_2$) are
real. Further, since $a_3 >0$ and the global minimum occurs at a
positive value, $\eta_2 > \eta_1 > 0$. Consequently, the last
inequality is satisfied when $\lambda < \eta_1$ or $\lambda >
\eta_2$ where \beq \eta_{1,2} = (a+b)^2 \pm \sqrt{(a^2-b^2)^2 + 4
k_{ie} k_{ei}}. \nonumber \eeq Note that $\eta_1$ is explicitly
seen to be positive by applying Eq. (\ref{stabo}). Further,
$\eta_2 > (a+b)^2
> 2 a b$. Thus the inequality $\lambda > \eta_2
> 2 a b$ is not possible given the stability condition $\lambda <
2 a b$ derived earlier. Therefore the last inequality in Eq.
(\ref{ineq2}) reduces to $\lambda < \eta_1$.

Summarizing, we get the following set of stability conditions:
\beq
\lambda & < & \frac{k_{ie} k_{ei}+a^2 b^2}{a b}, \nonumber \\
\lambda & < & 2 a b, \nonumber \\
\lambda & < & \eta_1 \nonumber. \eeq Let $\kappa=\min
\{\displaystyle{\frac{k_{ie} k_{ei}+a^2 b^2}{ab}}, 2 a b, \eta_1
\}$, then all these inequalities will be simultaneously satisfied
if \beq \lambda<\kappa . \label{stabsim2} \eeq

Thus the rightmost tip of the boundary curve along the real axis
is $(\kappa,0)$.  Therefore the distance function $\delta(G_{ii})$
is given by \beq \delta(G_{ii})=|\kappa-G_{ii}|, \quad i=1,2,
\ldots ,N. \eeq Applying Eq. (\ref{gnrlstbl}), we obtain the
following stability condition for the present model with symmetric
couplings:\beq \frac{1}{2}{\sum_{j \neq
i}}(|G_{ji}|+|G_{ij}|)+G_{ii} \leq \kappa, \quad i=1,2, \ldots ,
N. \label{modelBstbl} \eeq As we discussed before, this condition
is satisfied if the individual coupling strengths obey the
following stability constraints: \beq |c_{np}|<\kappa,\quad
\mbox{for} \quad c_{np}=c_{pn},\quad n, p =1,2, \ldots , N. \eeq

\section{Conclusions}

We have presented a general method for studying the stability of
the equilibrium state in neural network models. When the
single-neuron coupled networks, such as Hopfield type of models,
are studied, the stability result from our general approach
coincides with the known result found in the literature. As a
harder application, two typical neural population models where the
individual nodes are higher dimensional were considered. The
stability of the first model, a coupled network of two dimensional
systems, was solved completely.  For the second model, a coupled
network of four dimensional systems, stability criteria for
symmetric coupling was given analytically. For  the non symmetric
case, our method was used to obtain numerical criteria. Through
the above examples we have demonstrated that our general method is
applicable to arbitrary neural networks where the individual nodes
can themselves be high dimensional. When the dimension of the
individual node is not too high, analytical results are possible.

>From the stability criteria, we also derived simple bounds on the
coupling strengths which ensure stability. These bounds put a
limit on the magnitude of change that the coupling strengths can
undergo in the process of statistical learning.

\newpage
\section*{Appendix I}

In this Appendix, we state the Lienard-Chipart and generalized
Routh-Hurwitz criteria. The statements are taken directly from
Gantmacher \cite{gantmacher} and are given here for the sake of
completeness.

\subsection*{A. Lienard-Chipart Criterion}

Consider a real polynomial \beq f(z) = a_0 z^n + a_1 z^{n-1} +
\cdots + a_n, \nonumber \eeq with $a_0 > 0$. Necessary and
sufficient conditions for all the zeros of the polynomial to have
negative real parts can be given in any {\em one} of the following
forms \cite{gantmacher}:
\begin{enumerate}
\item $a_n > 0$, $a_{n-2} > 0$, $\ldots$; $\Delta_1 > 0$,
$\Delta_3 > 0$, $\ldots$, \item $a_n > 0$, $a_{n-2} > 0$,
$\ldots$; $\Delta_2 > 0$, $\Delta_4 > 0$, $\ldots$, \item $a_n >
0$, $a_{n-1} > 0$, $a_{n-3} > 0$, $\ldots$; $\Delta_1 > 0$,
$\Delta_3 > 0$, $\ldots$, \item $a_n > 0$, $a_{n-1} > 0$, $a_{n-3}
> 0$, $\ldots$; $\Delta_2 > 0$, $\Delta_4 > 0$, $\ldots$.
\end{enumerate}
Here $\Delta_p$ is the Hurwitz determinant of order $p$ given by
the formula \beq \Delta_p = \vline
\begin{array}{cccccccc}
a_1 & a_3 & a_5 & \ldots & & & & \\
a_0 & a_2 & a_4 & \ldots &  & & & \\
0 & a_1 & a_3 & \ldots &  & & & \\
0 & a_0 & a_2 & a_4 &  & & & \\
  &     &     &        & \cdot & & &  \\
  &     &     &        &       & \cdot & & \\
  &     &     &        &       & & \cdot & \\
  &     &     &        & & & & a_p
\end{array}
\vline \ , \ \ \ p=1,2, \ldots, n, \nonumber \eeq where $a_k=0$
for $k > n$. In the literature, the equivalent Routh-Hurwitz
criterion is usually used. But the Lienard-Chipart is better since
the number of determinants that have to be evaluated is half the
number that have to be evaluated for the Routh-Hurwitz criterion.
This leads to a simpler set of inequalities that need to be
evaluated. In the main text, we use the third form of the
Lienard-Chipart criterion given above.

\subsection*{B. Generalized Routh-Hurwitz Criterion}

Consider a polynomial $f(z)$ with complex coefficients. Suppose
that \beq f(i z) = b_0 z^n + b_1 z^{n-1} + \cdots b_n + i (a_0 z^n
+ a_1 z^{n-1} + \cdots + a_n), \nonumber \eeq where $a_0$, $a_1$,
$\ldots$, $a_n$, $b_0$, $b_1$, $\ldots$, $b_n$ are real numbers.
If the degree of $f(z)$ is $n$, then $b_0+ i a_0 \neq 0$. Without
loss of generality, we may assume that $a_0 \neq 0$. Otherwise, we
consider the polynomial $g(z) = -i f(z)$ and repeat the analysis
for this polynomial. Both $f(z)$ and $g(z)$ have the same set of
zeros and so no information is lost. This is the case considered
in the main text.

If $\nabla_{2n} \neq 0$, then all the zeros of $f(z)$ have
negative real parts if \beq \nabla_2 > 0, \ \ \nabla_4 > 0, \ \
\ldots , \ \ \nabla_{2n} > 0, \nonumber \eeq where \beq
\nabla_{2p} = \vline
\begin{array}{cccc}
a_0 & a_1 & \ldots & a_{2p-1} \\
b_0 & b_1 & \ldots & b_{2p-1} \\
0 &  a_0 & \ldots & a_{2p-2} \\
0 &  b_0 & \ldots & b_{2p-2} \\
\ldots  &  \ldots   &  \ldots & \ldots
\end{array}
\vline \ , \ \ \ p=1,2, \ldots, n, \nonumber \eeq where
$a_k=b_k=0$ for $k>n$. Note that the condition $\nabla_{2n} \neq
0$ would be satisfied for a generic set of parameter values. This
is especially true in our case where $a_k$, $b_k$ are functions of
system parameters.

\section*{Appendix II}

The distance $\gamma$ from the center $(G_{ii},0)$ of the $i$th
Gershg\"{o}rin disc to any point on the boundary of the stability
zone is given by \beq \label{dist}
\gamma^2=(\lambda_R-G_{ii})^2+\lambda_I^2 \nonumber \eeq
Substituting $\lambda_I$ from Eq. (\ref{bdryeqs}) and
differentiating with respect to $\lambda_I$, we have \beq
\lefteqn{\frac{d\gamma^2}{d\lambda_R} = } \nonumber \cr &&
2(\lambda_R-G_{ii})-
\frac{(a+b-\lambda_R)^2}{(\lambda_R-a)}+\frac{[(\lambda_R-a)^2-b^2]
(ab+k_{ie}k_{ei}-b\lambda_R)}{b(\lambda_R-a)^2}. \nonumber \eeq

Setting $\displaystyle{\frac{d\gamma^2}{d\lambda_R}}=0$, we get
two solutions:\beq \lambda_R=a \pm
b\sqrt{\frac{k_{ie}k_{ei}}{2b(a+b-G_{ii}+k_{ie}k_{ei})}}.
\nonumber \eeq Since the boundary of $\Omega$ lies to the right of
the point (a,0), we can discard the smaller solution. Substituting
the remaining solution in the equation for $\gamma^2$ and taking
the square root, we get the shortest distance as: \beq
\lefteqn{\delta_i=\gamma_{min}  = } \nonumber \cr &&
\sqrt{(a-G_{ii})^2-b^2-2k_{ie}k_{ei}+2\sqrt{k_{ie}k_{ei}[2b(a+b-G_{ii})+k_{ie}k_{ei}]}},\cr
&& \quad i=1,2, \ldots ,N. \nonumber \eeq

\section*{Acknowledgments}

The work was supported by US ONR Grant N00014-99-1-0062. GR is
also supported by the Department of Science and Technology, India
and the Homi Bhabha Fellowship.

\newpage
\section*{Figure Caption}

\begin{description}
\item{\bf Figure 1:} Schematic of the stability zone. \item{\bf
Figure 2:} Schematic of the network configuration. \item{\bf
Figure 3:} Stability zone for model Eq.(\ref{neteqns})
 \item{\bf
Figure 4:} Stability zone for model Eq.(\ref{neteqns2})
\end{description}

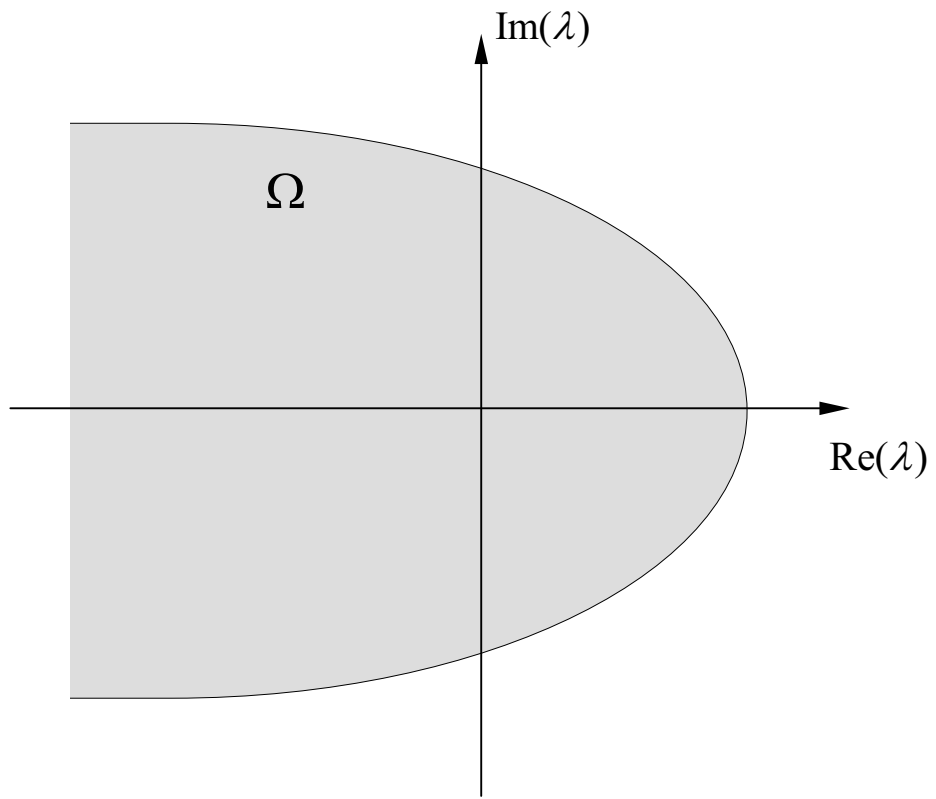
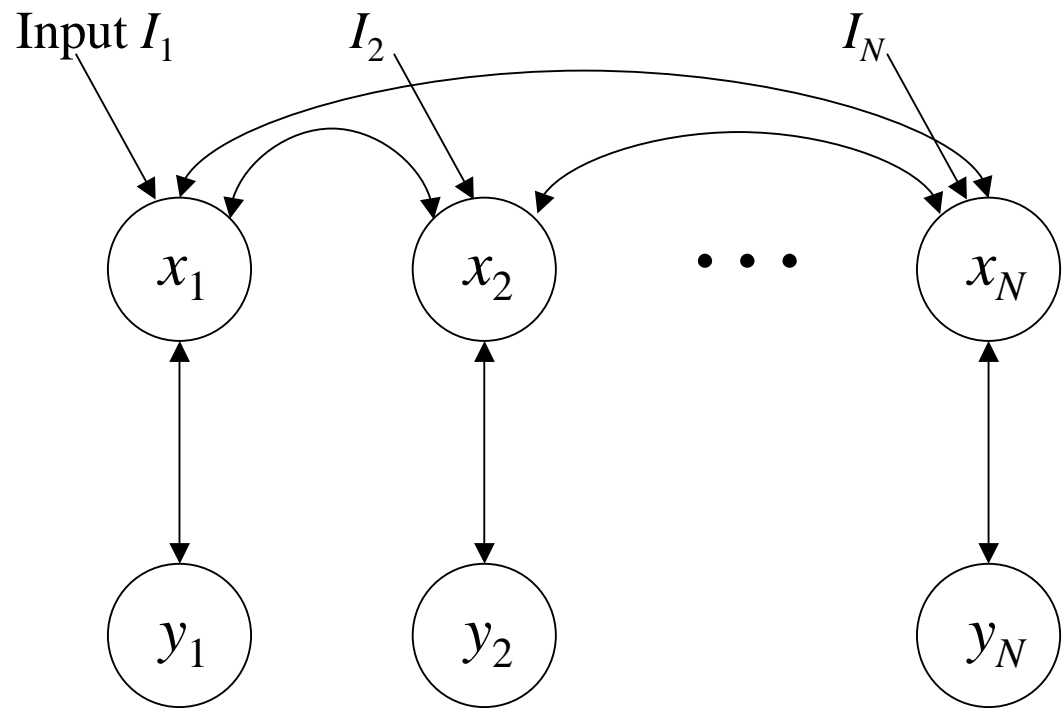
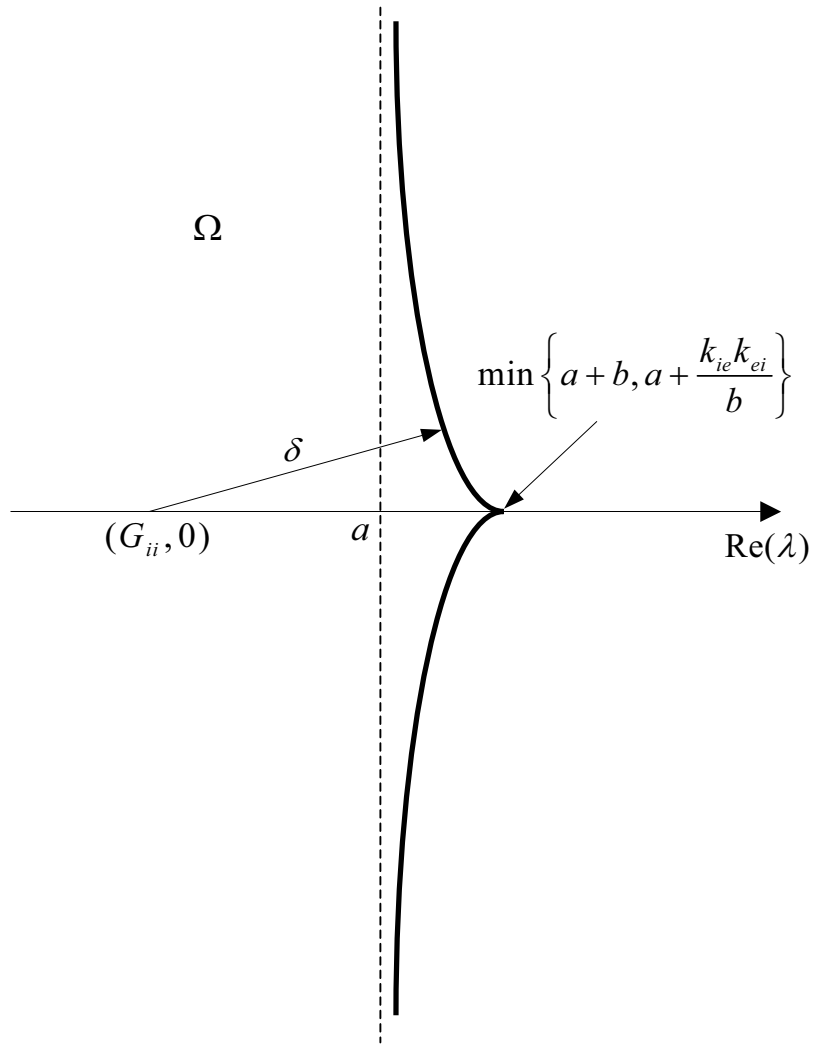
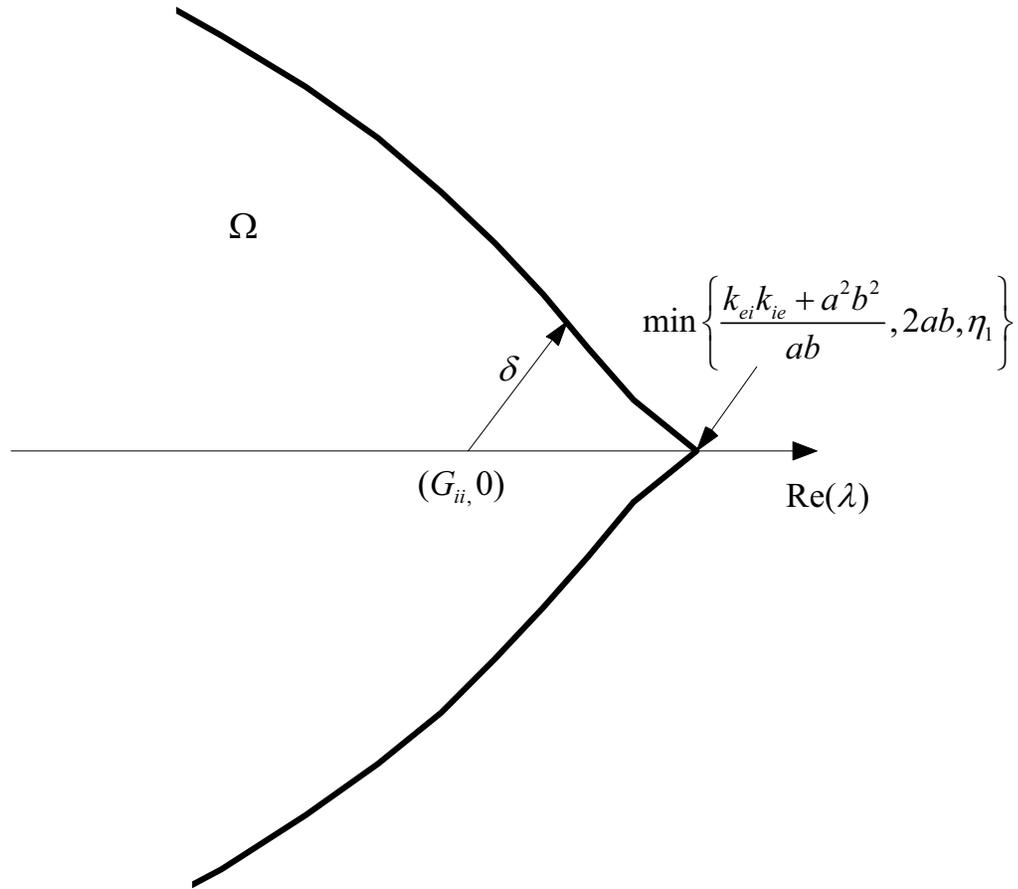
\end{document}